\input{aipcheck}

\documentclass[
    ,final            
  ]
  {aipproc}

\layoutstyle{8x11double}
\def\sqr#1#2{{\vcenter{\hrule height.#2pt
      \hbox{\vrule width.#2pt height#1pt \kern#1pt
          \vrule width.#2pt}
      \hrule height.#2pt}}}
\def\square{\mathchoice\sqr34\sqr34\sqr{2.1}3\sqr{1.5}3}

\begin{document}

\title{Planetary Nebula Studies of Face-On Spiral Galaxies:  Is the Disk Mass-to-Light Ratio Constant?}


\author{Kimberly A. Herrmann}{
  address={Department of Astronomy and Astrophysics, 
Pennsylvania State University}
}

\author{Robin Ciardullo}{
  address={Department of Astronomy and Astrophysics, 
Pennsylvania State University} 
}

\begin{abstract}
When astronomers study the dark matter halos of spiral galaxies, they normally assume that the disk mass-to-light ratio is {\it constant\/}.  We describe a method of analyzing the kinematics of planetary nebulae (PNe) in nearby face-on spiral galaxies to test this assumption.  Since the restoring force for stellar motions perpendicular to the galactic disk is proportional to the disk mass surface density, measurements of the vertical velocity dispersion can be used to produce an independent measure of the total amount of matter in the disk.  Our steps are: (1) to identify a population of PNe by imaging the host spiral in several filters, and (2) to isolate the vertical velocity dispersion from spectroscopic observations of the PNe.  Our first results for the PNe of M33 indicate that the mass-to-light ratio of the galaxy's disk actually {\it increases\/} by more than a factor of 5 over the inner 6 disk scale lengths.  We have begun similar studies of the PNe in five more face-on galaxies: M83, M101, M94, NGC 6946, and M74.  These data will also produce additional science such as galaxy distances and constraints on the disk transparency.
\end{abstract}
\maketitle
\section{Introduction and Scientific Motivation}
Dark matter, a mysterious topic, has been under intense study for many years.  One way to shed some light on the subject is to examine galactic halos.  In the case of spiral galaxies with weak bulges, one typically measures the total galactic mass via the system's rotation curve (Faber \& Gallagher 1979; Ashman 1992; Combes 2002), assumes that the mass near the center of the galaxy is entirely baryonic, and subtracts off the contribution of the disk by assuming a constant disk mass-to-light ratio (Kent 1986, Palunas \& Williams 2000; Sofue et al.\ 2003).  (This is the ``maximal disk'' method.)  However, while absorption line studies (Bottema 1993; Bottema, van der Kruit, \& Freeman 1987; van der Kruit et al.\ 2001; Gerssen, Kuijken, \& Merrifield 1997, 2000) have indicated that the disk mass-to-light ratio is constant in the {\it inner\/} regions of spiral galaxies, there formerly were no results about the {\it outer\/} regions where the influence of dark matter is greatest.  The extreme difficulty in separating the mass of a galaxy's visible disk from 
that of its dark halo limits our understanding of almost every facet of galaxy formation.  An independent method of determining the disk mass is needed to break the disk-halo degeneracy.
\section{Our Method: Using Planetary Nebulae to study Disk Mass}

To determine a disk's mass-to-light ratio, one can study the vertical motions of old disk stars.  Since the restoring force for stellar motions perpendicular to a galactic disk is proportional to the disk mass surface 
density ($\Sigma$), measurements of the vertical velocity dispersion ($\sigma_z$) immediately yield an independent measure of the amount of matter in a disk.  From the isothermal disk approximation, old disk stars oscillate in $z$ according to
\begin{equation}
\sigma_z^2(R) = \pi G\Sigma(R) z_0,
\end{equation}
\noindent
where $z_0$ is the scale height of the stars (Binney \& Tremaine 1987).  Since studies of edge-on spirals demonstrate that $z_0$ is constant with radius (van der Kruit \& Searle 1981; Bizyaev \& Mitronova 2002), this parameter can be fixed at some appropriate value.  Then, by observing a face-on galaxy, we can see if the matter scale length does indeed decline in a manner similar to the light.

The traditional method of determining $\sigma_z$ via absorption line spectroscopy is very challenging both observationally and in terms of data analysis.  Moreover, because the technique is limited by surface brightness, its effectiveness is restricted to a galaxy's inner regions (Bottema 1993; Gerssen, Kuijken, \& Merrifield 1997, 2000).  An alternate method to measure $\sigma_z$ is to use planetary nebulae (PNe) as kinematic test particles.  PNe are relatively numerous, easy to detect in a galaxy's outer regions where dark matter is most important, and come from a progenitor population of low and intermediate mass stars.  They are therefore representative of the old stellar disk of a galaxy.  In addition, since PNe are strong emission line sources, virtually every object that can be found photometrically can be observed spectroscopically.  With a medium ($R \sim 5000$) resolution instrument, radial velocity measurements to a precision of $\sim 2$ km s$^{-1}$ are obtainable without much difficulty.  This makes PNe ideal test particles for probing the disk mass of face-on spirals.
\section{Finding the PNe}
In order to study the disk mass-to-light ratio, we must first identify a suitably large population of PNe.  The spiral galaxy under study must be imaged with a 4-m class telescope in four filters: two narrow on-band
filters centered at [O~III] $\lambda 5007$ and H$\alpha$ (in the rest-frame of the galaxy), and two wider off-band filters, such as $V$ and $R$.  Since PNe have virtually no continuum, they can be found by blinking the on-band images against their off-band counterparts.  True PN candidates are: (1) consistent with point sources, (2) detected in $\lambda 5007$ but invisible in $V$ and $R$, and (3) invisible or weak in H$\alpha$.  These criteria exclude most or all H~II regions and supernova remnants.  At this stage, we can determine the distance to the galaxy via the Planetary Nebula Luminosity Function (Ciardullo et al.\ 2002), and attempt to constrain the transparency of the disk via PN number counts.  Thus far, as part of our 
program, we have discovered 152 PNe in the Triangulum Galaxy, M33 (Ciardullo et al.\ 2004; hereafter C04), 65~PNe in the Northern Pinwheel Galaxy, M101 (Feldmeier, Ciardullo, \& Jacoby 1996), and $\sim 200$ PNe in the Southern Pinwheel Galaxy, M83 (Herrmann, Ciardullo, \& Vinciguerra 2005).


\begin{figure}[t]
\includegraphics[height=4.94in]{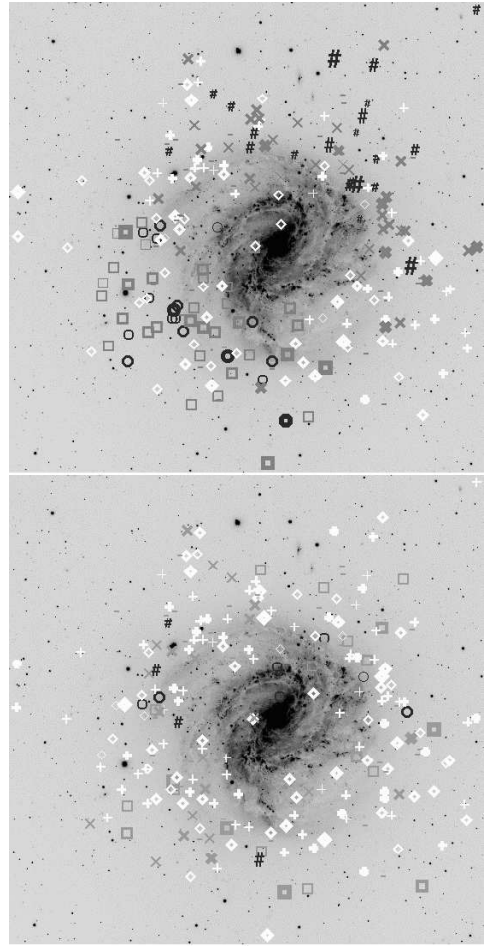}
\caption{PN radial velocities in M83.  From fastest away to fastest toward the symbols are $\circ$, $\square$, $\diamond$, $+$, $\times$, \#.  PNe not observed are indicated with a dash.  The top panel shows the radial velocities as observed; the bottom illustrates the velocities after the removal of 
galactic rotation.}
\end{figure}
\section{PNe Kinematics}
Once we have identified a large population of PNe and determined accurate positions, the next step is to obtain a high precision radial velocity for each object.  This can be done with fiber-coupled spectrographs, such as the Hydra instruments on the WIYN and Blanco 4-m telescopes, and the Medium Resolution Spectrograph on the Hobby-Eberly Telescope.  In order to minimize systematic errors, it is useful to target the PNe multiple times and to pay special attention to the wavelength calibration.  Ideally, the velocity 
uncertainties should be $< 5$ km s$^{-1}$.

The top frame of Fig.~1 shows the preliminary radial velocities of 203 PNe in M83 after correcting for the systemic and barycentric velocities.  PNe with the largest velocities {\it away\/} from us are clustered in the lower left of the frame, while PNe with the largest velocities {\it toward\/} us are in the upper right.  Clearly, we are detecting the rotation of the galaxy.  For a flat axisymmetric system, the radial velocity, $v_{rad}$ is given by
\begin{equation}
v_{rad} = v_{\phi}\cos\phi\sin i + v_R\sin\phi\sin i + v_z \cos i,
\end{equation}
\noindent
where $v_{\phi}$, $v_R$ and $v_z$ are the azimuthal, radial and vertical components, $i$ is the inclination, and $\phi$ is the angle from the principal axis in the galaxy plane.  The rotation component is eliminated by subtracting out $v_{\rm rot}\cos\phi\sin i$ where $v_{\rm rot}$ is taken from studies of H~I gas (Tilanus \& Allen 1993).  The resulting residual velocities are random with position in the galaxy, indicating that the rotation has been taken out.  (See the bottom frame of Fig.~1.)
\begin{figure}[t]
\includegraphics[height=2.0in]{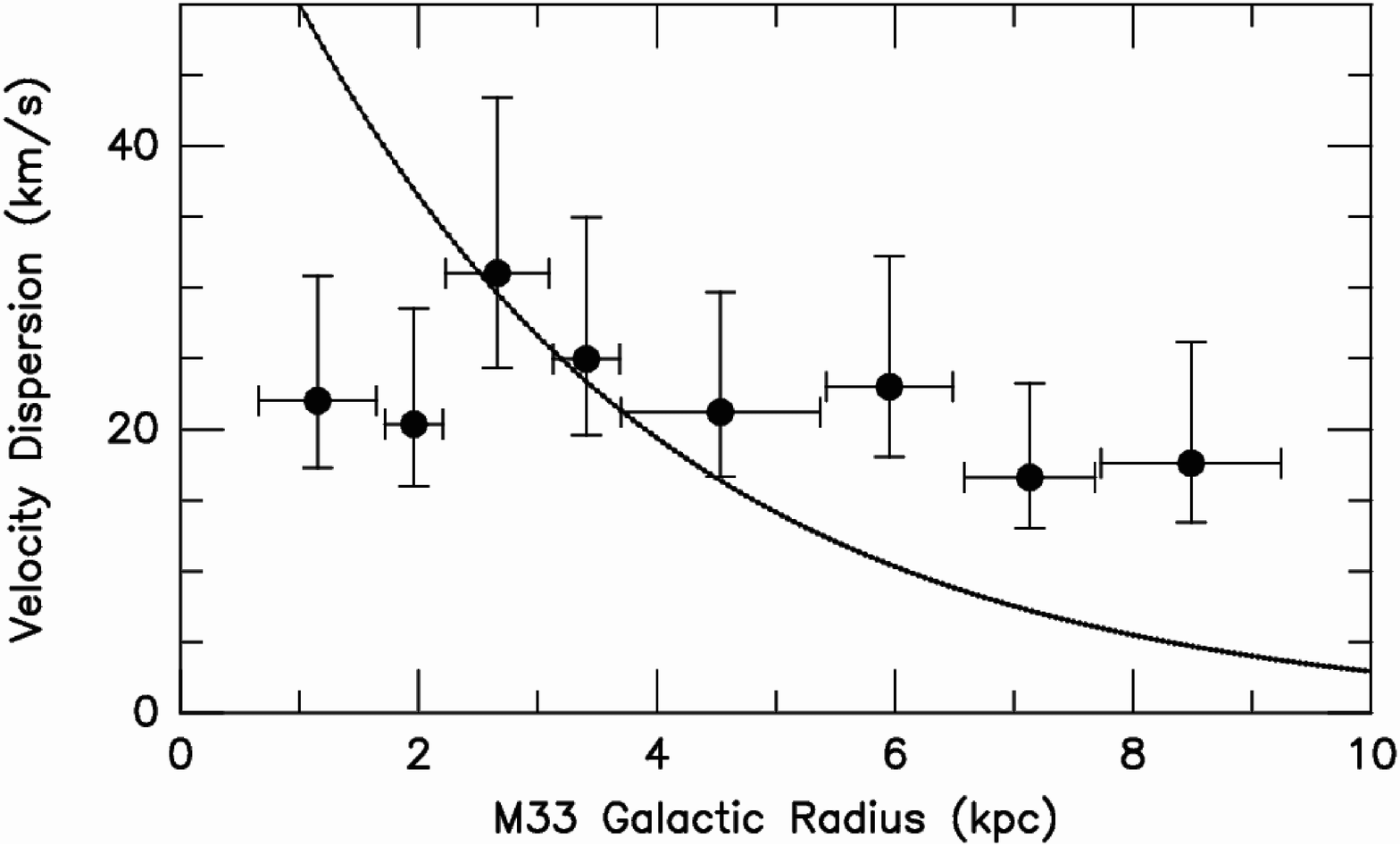}
\includegraphics[height=2.0in]{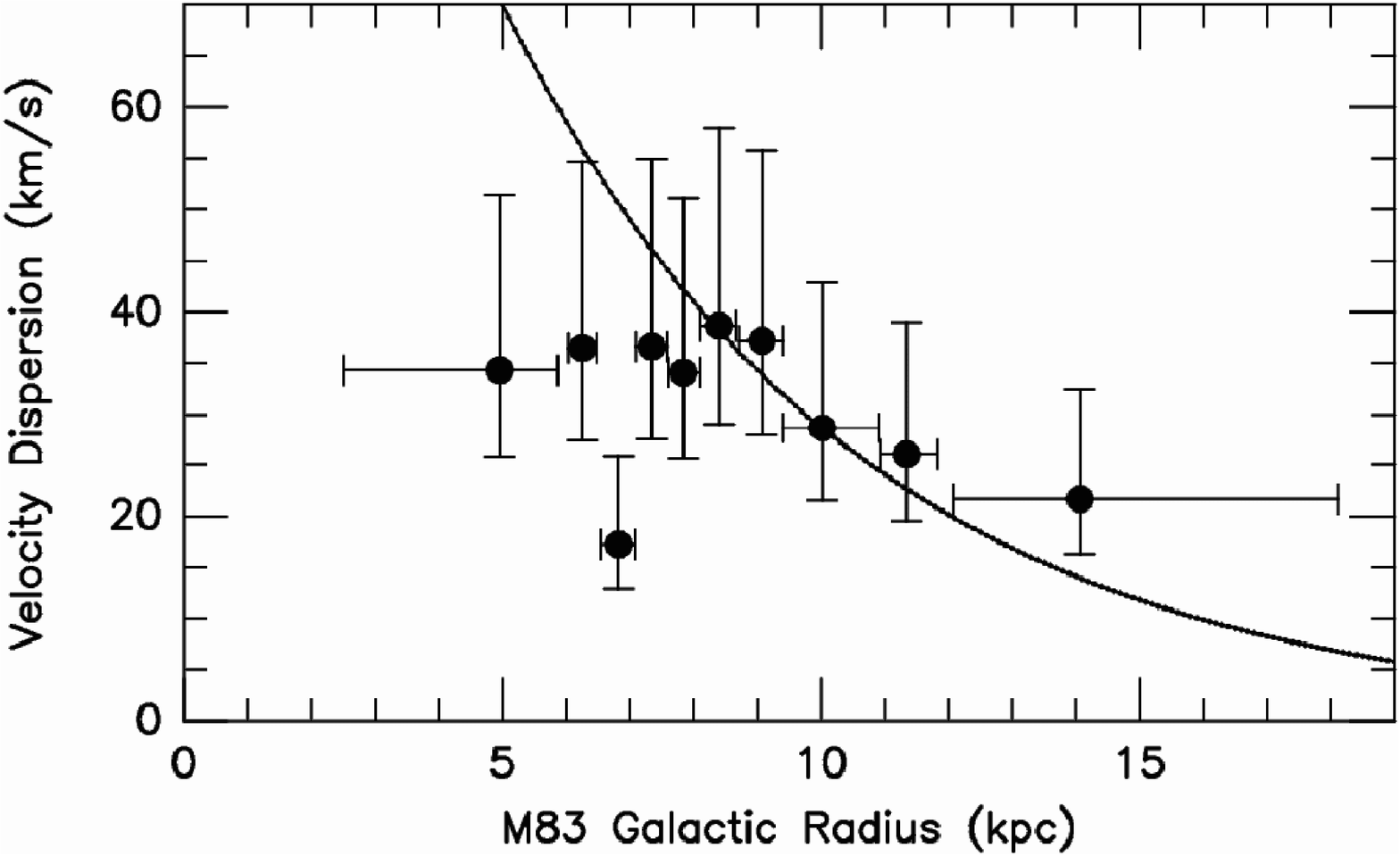}
\caption{Dispersion in the residual PN radial velocities for M33 and M83 after galactic rotation has been removed.  The curves show the exponential decay expected for constant mass-to-light ratio disks.  The error bars in $y$ represent 90\% confidence intervals, while those in $x$ indicate the bin size.}
\end{figure}

If the line-of-sight velocity measurements are dominated by the vertical component of the velocity ellipsoid, then the dispersion in the residual velocities, $\sigma_{rad}$, should be related to the disk mass surface density, $\Sigma$, via Eqn.~(1).  For a constant mass-to-light ratio disk, this implies that $\sigma_{rad}$ should fall off exponentially with a scale-length twice that of the galaxy's light.  Figure~2 shows the line-of-sight residual velocity dispersion for the PNe in M33 and M83.  For M33, the dispersion clearly does not follow the light.  This is not entirely unexpected, since the galaxy's $\sim 56^\circ$ inclination ensures that all three components of the velocity ellipsoid contribute to $\sigma_{rad}$.  Our preliminary results for M83 are similar.  Thus, in both cases, we need to de-couple $\sigma_z$ from the other components of stellar
motion.

The observed line-of-sight velocity dispersion of a galaxy can be written in terms of the azimuthal ($\phi$), radial ($R$), and perpendicular ($z$) velocity dispersions via
\begin{equation}
\sigma_{rad}^2 = \sigma_{\phi}^2\cos^2\!\phi\sin^2\!i + 
\sigma_R^2\sin^2\!\phi\sin^2\! i + \sigma_z^2\cos^2\!i + \sigma_{meas}^2,
\end{equation}
where $\sigma_{meas}$ is the measurement uncertainty.  (Note that to extract $\sigma_z$ from $\sigma_{rad}$, $\sigma_{meas}$ must be kept to a minimum.) Since there are three unknowns but only one equation, external constraints are needed.  One such constraint is the epicyclic approximation, which allows us to remove $\sigma_{\phi}$ from the equation by writing it in terms of $\sigma_R$ and the radial gradient of the circular velocity (Binney \& Tremaine 1987).  Two others are the Toomre (1964) criterion, which requires that the disk be stable against axisymmetric perturbations, and the firehose instability, which forces us to consider only those disks that are stable against buckling (Toomre 1966; Merritt \& Sellwood 1994).  Finally, for non-barred galaxies, we can impose the Morosov (1980, 1981a, 1981b) criterion, 
which requires that a disk be stable against the formation of a bar.

Each of these constraints eliminates some combination of $\sigma_z$ and $\sigma_R$ from consideration.  We can then use a maximum-likelihood analysis to determine which of the remaining combinations of the two variables are most probable.  If enough PNe are observed, we can also determine the system's asymmetric drift, and place a further constraint on the solution (Binney \& Tremaine 1987).
\section{Results for M33}
At present, M33 is the only galaxy for which we have a complete analysis.  The most likely values of $\sigma_z$ and $\sigma_R$ are shown in Fig.~3.  The data demonstrate why the line-of-sight velocity dispersion of M33 varies  so little with radius.  Near the center of the galaxy, the increase in $\sigma_z$ is negated by a turnover in the radial velocity dispersion.  (If this did not happen, then $\sigma_R$ in the central $\sim 1$~kpc would be greater than $v_{\rm rot}$, and the galaxy would have a bulge.)  At larger
radii, both $\sigma_R$ and $\sigma_z$ decline exponentially, with the scale length of $\sigma_R$ being $\sim 25\%$ larger than that of $\sigma_z$.  The derived values of the dispersion ratio, as well as its radial gradient, are in excellent agreement with models of disk heating (Villumsen 1985; Jenkins \& Binney 1990; Carlberg 1987).
\begin{figure}[b!]
\includegraphics[width=3.1in,height=3.1in]{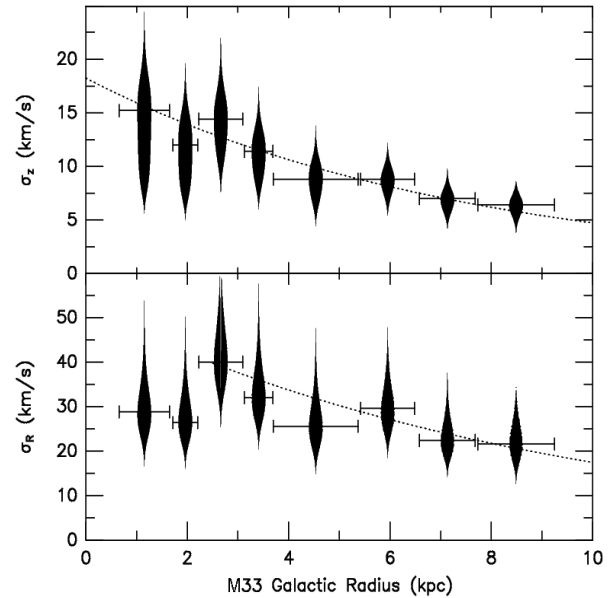}
\caption{The vertical and radial dispersions of M33 PNe derived from a maximum-likelihood decomposition of $\sigma_{rad}$.  The non-Gaussian uncertainties are illustrated by the width of the lines.   The dotted curves show the best-fit exponentials.}
\end{figure}

A more surprising result is that the mass scale length inferred from $\sigma_z$ is more than twice that expected from observations of the galaxy's infrared light.  Near the center of the galaxy, the disk mass-to-light ratio of $M/L_V \sim 0.3$ agrees with that derived from the galaxy's rotation curve.  However, at 9~kpc (6 $K$-band disk scale lengths), $M/L_V \sim 1.5$ (see Fig.~4).  Since this increase runs counter to inside-out scenarios of galaxy formation, we must look for possible errors in our analysis.
\begin{figure}[t]
\includegraphics[height=1.82in]{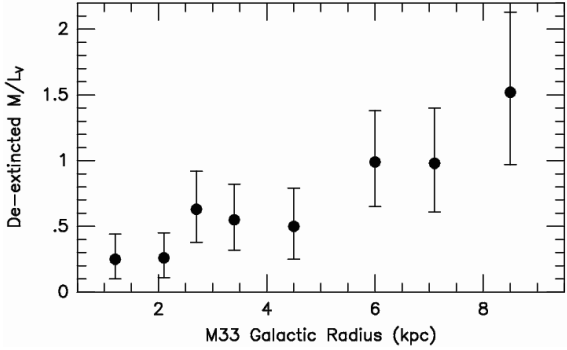}
\caption{The mass-to-light ratio of M33's disk as a function of galactic radius.  The error bars do not include systematic errors.}
\end{figure}
\section{Possible Problems}
There are several possible sources of error in our analysis.  Some H~II regions may be contaminating the PN sample and artificially lowering the PN velocity dispersion. This is unlikely, since our photometric procedures should have eliminated virtually all such objects, but it must be considered.  Similarly,
though our analysis used an assumed value of 175~pc for the PN scale height, a 25\% change in this number does not significantly alter the results.  Finally, our measurement of the disk matter scale length could
be affected by a systematic change in the galaxy's internal extinction, or a breakdown in the various stability arguments.  Our analysis suggests that such a breakdown would cause us to underestimate the disk scale length, not overestimate it, but this possibility cannot be excluded (see C04).  Further observations of more face-on galaxies are needed to check these assumptions.
\section{Conclusions and Future Work}
For at least two decades, astronomers have been studying dark matter halos by assuming that disk mass-to-light ratios are constant.  Our results for the PN kinematics of M33 suggests that this assumption is not valid.  However, because M33 is inclined $56^\circ$ to the line-of-sight, there are a number of caveats associated with our conclusions.  To remove these uncertainties, we have begun studies of five additional galaxies which are more face-on than M33, including  M83, M101, M94, NGC 6946, and M74.  These 
data should tell us whether the spiral disks are truly super-maximal, as the M33 data suggests, or if $\sigma_R$ and $\sigma_{\phi}$ are conspiring against us.  


\begin{theacknowledgments}
We would like to thank the meeting organizers for all their hard work on arranging an excellent meeting.  This work was supported by NSF grant AST 00-71238, a Zaccheus Daniel travel grant, a Pennsylvania NASA Space Grant, and by CTIO travel support.
\end{theacknowledgments}

\bibliographystyle{aipproc}

\begin{thebibliography}{22}

\bibitem{1}
Ashman, K.M. 1992, \emph{Pub.~A.S.P.}, \textbf{104}, 1109

\bibitem{2}
Binney, J., \&  Tremaine, S. 1987, \emph{Galactic Dynamics}, (Princeton:
Princeton Univ.~Press)

\bibitem{3}
Bizyaev, D., \& Mitronova, S. 2002, \emph{Astr.~Ap.}, \textbf{389}, 795

\bibitem{4}
Bottema, R. 1993, \emph{Astr.~Ap.}, \textbf{275}, 16

\bibitem{5}
Bottema, R., van der Kruit, P.C., \&  Freeman, K.C. 1987, \emph{Astr.~Ap.}, 
\textbf{178}, 77

\bibitem{6} 
Carlberg, R.G. 1987, \emph{Ap.~J.},  \textbf{322}, 59

\bibitem{7}
Ciardullo, R., Durrell, P.R., Laychak, M.B., Herrmann, K.A., Moody, K., 
Jacoby, G.H., \& Feldmeier, J.J. 2004, \emph{Ap.~J.}, \textbf{614}, 167

\bibitem{8}
Ciardullo, R., Feldmeier, J.J., Jacoby, G.H., de Naray, R.K., Laychak, M.B.,
\& Durrell, P.R. 2002, \emph{Ap.~J.}, \textbf{577}, 31

\bibitem{9}
Combes, F. 2002, \emph{New Astronomy Rev.}, \textbf{46}, 755

\bibitem{10}
Faber, S.M., \& Gallagher, J.S. 1979, \emph{ARA\&A}, \textbf{17}, 135

\bibitem{11}
Feldmeier, J.J., Ciardullo, R., \& Jacoby, G.H. 1996, \emph{Ap.~J.}, 
\textbf{461}, L25

\bibitem{12}
Gerssen, J., Kuijken, K., \& Merrifield, M.R. 1997, \emph{MNRAS}, \textbf{288}, 618

\bibitem{13}
Gerssen, J., Kuijken, K., \& Merrifield, M.R. 2000, \emph{MNRAS}, \textbf{317},
545

\bibitem{14}
Herrmann, K.A., Ciardullo, R., \& Vinciguerra, M. 2005, \emph{Bull.~A.A.S.}, 
\textbf{205}, 138.12 (2005).

\bibitem{15} 
Jenkins, A., \& Binney, J. 1990, \emph{MNRAS}, \textbf{245}, 305

\bibitem{16}
Kent, S.M. 1986, \emph{A.J.}, \textbf{91}, 1301

\bibitem{17} 
Merritt, D., \& Sellwood, J.A. 1994, \emph{Ap.~J.}, \textbf{425}, 567

\bibitem{18}
Morosov, A.G. 1980, \emph{Soviet Astron.}, \textbf{24}, 391

\bibitem{19}
Morosov, A.G. 1981a, \emph{Soviet Astron.}, \textbf{25}, 19

\bibitem{20}
Morosov, A.G. 1981b, \emph{Soviet Astron.}, \textbf{25}, 421

\bibitem{21}
Palunas, P., \&  Williams, T.B. 2000, \emph{A.J.}, \textbf{120}, 2884

\bibitem{22}
Sofue, Y., Koda, J., Nakanishi, H., \& Onodera, S. 2003, \emph{Pub.~A.S.J.}, 
\textbf{55}, 59

\bibitem{23}
Tilanus, R.P.J., \& Allen, R.J.\ 1993, \emph{Astr.~Ap.}, \textbf{274,} 707 

\bibitem{24}
Toomre, A. 1964, \emph{Ap.J.}, \textbf{139}, 1217

\bibitem{25} Toomre, A. 1966, in \emph{Notes on the 1966 Summer Study 
Program in Geophysical Fluid Dynamics at the Woods Hole Oceanographic
Institution}, (Woods Hole: Woods Hole Oceanographic Institute), 111

\bibitem{26}
van der Kruit, P.C., Jim\'enez-Vicente, J., Kregel, M., \& Freeman, K.C.
2001, \emph{Astr.~Ap.}, \textbf{379}, 374

\bibitem{27}
van der Kruit, P.C., \& Searle, L. 1981, \emph{Astr.~Ap.}, \textbf{95}, 105

\bibitem{28} 
Villumsen, J.V. 1985, \emph{Ap.~J.}, \textbf{290}, 75

\end{thebibliography}

\end{document}